# PERIODICITIES IN SOLAR NEUTRINO FLUX DATA FROM SAGE AND GALLEX-GNO DETECTORS

**Koushik Ghosh & Probhas Raychaudhuri**
**Department of Applied Mathematics**
**University of Calcutta**
**92, A.P.C. Road, Calcutta – 700 009**

**ABSTRACT:**

We have used both the Date-Compensated Discrete Courier Transform and Periodogram analysis of the monthly solar neutrino flux data from (1) SAGE detector during the period from 1$^{st}$ January 1990 to 31$^{st}$ December 2000; (2) SAGE detector during the period from April 1998 to December 2001; (3) GALLEX detector during the period from May 1991 to January 1997; (4) GNO detector during the period from May 1998 to December 2001; (5) GALLEX-GNO detector (Combined data) from May 1991 to December 2001 and (6) average of the data from GNO and SAGE detectors during the period from May 1998 to December 2001. (1) exhibits periodicity around 19, 24 and 80-85 months. (2) shows periodicity around 1.5 months. For (3) we observe periodicity around 1.6, 2.5 and 19 months. For (4) periodicity is seen around 1.4 and 3.4 months. (5) gives periodicity around 1.7 and 19 months while (6) shows periodicity around 1.8 and 2.2 months. We have found almost similar periods in the solar flares, sunspot data, solar proton data which indicates that the solar activity cycle is due to the pulsating character of nuclear energy generation inside the sun.

## I. INTRODUCTION:

Solar neutrino flux detection is very important towards the understanding of solar internal structure (i.e. its nuclear energy generation process, temperature, density etc.). Standard solar model (SSM) do not predict solar neutrino flux variation [1]. Only perturbed solar model of Raychaudhuri [2] suggested that the solar neutrino flux should vary with the solar activity cycle. Solar neutrino flux data from Homestake detector varies with the solar activity cycle [3,4,5,6] but at present it appears that there is no significant anticorrelation of solar neutrino flux data with the sunspot numbers. Raychaudhuri [7,8] showed that solar neutrino flux data from Kamiokande, Superkamiokande, GALLEX and SAGE are varying with the solar activity cycle. From the analysis it appears that variation of solar neutrino flux data indicates that there must be periodicity in the data.

Many authors analysed the solar neutrino flux data from Homestake detector and have found short-time periodicities around 5, 10, 15, 20, 25 months etc. [9, 10, 11, 12]. Raychaudhuri analysed the solar neutrino flux data from SAGE, GALLEX, Superkamiokande and have found that solar neutrino flux data varies with the solar activity cycle and have found periodicities around 5 and 10 months.

In this present paper two different methods are employed to find out the periodicity of the solar neutrino flux data obtained from the SAGE, GALLEX and GNO detectors. The obtained data is used for the analysis of periodicity following the methods 1) Ferraz-Mello [13] method of Data-compensated Discrete Fourier Transform and 2) Scargle [14] method of periodogram. For all these methods graphs are plotted to depict the periodic nature of the solar neutrino flux data and the corresponding periodicities are calculated.

## II. CALCULATION OF PERIODICITIES:

**1. Ferraz-Mello method of Date-Compensated Discrete Fourier Transform:**

The technique called Date-Compensated Discrete Courier Transform (DCDFT) corresponds to a curve-fitting approach using a sinusoid-plus-constant model and is summarised below. For each trial frequency $\omega$, one coefficient of spectral correlation S is obtained by the following formulae [13]:

$$a_o^{-2} = N \tag{1}$$

$$a_1^{-2} = \sum \cos^2 x_i - a_o^2 (\sum \cos x_i)^2 \tag{2}$$

$$a_2^{-2} = \sum \sin^2 x_i - a_o^2 (\sum \sin x_i)^2 - a_1^2 M^2 \tag{3}$$

where
$$M = \sum \cos x_i \sin x_i - a_o^2 (\sum \sin x_i)(\sum \cos x_i) \tag{4}$$

and,

$$c_1 = a_1 \sum f_i \cos x_i \tag{5}$$

$$c_2 = a_2 \sum f_i \sin x_i - a_1 a_2 c_1 M \tag{6}$$

$$S = \frac{c_1^2 + c_2^2}{\sum f_i^2} \tag{7}$$

N is the number of observations in the series, $t_i$ are the observation dates, $x_i = 2\pi\omega t_i$ and $f_i$ are the measures of data referred to the mean i.e. $f_i = y_i - \bar{y}$ so that $\sum f_i = 0$, where $y_i$'s are the observed data. All other symbols are intrinsic quantities. The summations are made for i=1 to i=N. Usually the range of frequencies is considered from the observed frequency $\omega_{obs}=1/N$ to the Nyquist frequency $\omega_N$.

To decide whether the peaks in the graph are significant or not we use the test derived by G.R. Quast [15] which is given by the following expressions:

$$G = -\frac{N-3}{2} \ln(1-S) \tag{8}$$

$$H = \frac{N-4}{N-3}(G + e^{-G} - 1) \tag{9}$$

$$\alpha = \frac{2(N-3)\Delta t \cdot \Delta\omega}{3(N-4)} \tag{10}$$

$$C = (1 - e^{-H})^\alpha \tag{11}$$

where $\Delta t$ is the time interval covered by the observations and $\Delta\omega$ is the range of frequencies sampled. C is the confidence of the result; $(1 - C)$ may be interpreted as the probability of having the height of the highest peak by chance only. Here we consider a suitable number of values of $\omega$ from $\omega_{obs} = 1/N$ to $\omega_N$ and we arrange the corresponding months from $1/\omega_N$ to $1/\omega_{obs}$ at equal intervals and we obtain corresponding magnitudes of H and C at those months.

2.  **Scargle method of periodogram:**

For a time series $X(t_i)$, where i = 1, 2, …, N the periodogram as a function of the frequency $\omega$ is defined as [14]:

$$P_X(\omega) = \frac{1}{2} = \left\{ \frac{[\sum_{i=1}^{N} X(t_i) \cos\omega(t_i - \tau)]^2}{\sum_{i=1}^{N} \cos^2\omega(t_i - \tau)} + \frac{[\sum_{i=1}^{N} X(t_i) \sin\omega(t_i - \tau)]^2}{\sum_{i=1}^{N} \sin^2\omega(t_i - \tau)} \right\} \tag{12}$$

where $\tau$ is defined by the equation

$$\tan(2\omega\tau) = \left(\sum_{i=1}^{N} \sin 2\omega t_i\right) / \left(\sum_{i=1}^{N} \cos 2\omega t_i\right) \tag{13}$$

Here also we consider 1000 values of $\omega$ from $\omega_{obs} = 1/N$ to $\omega_N$ and we arrange the corresponding months from $1/\omega_{NM}$ to $1/\omega_{obs}$ at equal intervals and we obtain corresponding $P_X(\omega)$ at those months.

## III. RESULTS:

| Data | Periods (in months) | |
|---|---|---|
| | Ferraz-Mello method of Date-Compensated Discrete Fourier Transform [the corresponding levels of confidence are given in the brackets] | Scargle Method of Periodogram |
| (1) SAGE data (1st January 1990 to 31st December 2000) | 19.007 (96.01%), 23.720 (99.95%), 80.413 (99.37%) | 18.614, 23.851, 48.859, 85.258 |
| (2) SAGE date (April 1998 to December 2001) | 1.501 (93.52%), 14.767 (96.76%) | 1.504, 2.993, 4.584, 21.872 |
| (3) GALLEX data (May 1991 to January 1997) | 1.647 (99.38%), 2.541 (9.34%), 18.890 (91.23%) | 1.647, 2.541, 18.635, 46.990 |
| (4) GNO data (May 1998 to December 2001) | 1.403 (98.55%), 3.399 (99.62%) | 1.403, 3.433 |
| (5) Combined GALLEX-GNO data (May 1991 to December 2001) | 1.694 (96.23%), 19.002 (93.73%) | 1.694, 5.057, 18.507, 28.496, 41.946, 69.638 |
| (6) Average of the GNO and SAGE data (May 1998 to December 2001) | 1.843 (97.88%), 2.185 (97.74%) | 1.843, 2.185 |

## IV. DISCUSSION

Comparing the results obtained in Ferraz-Mello method and Scargle method we can confirmly say that (1) shows periodicities around 19 months, 23.7-23.9 and 80.4-85.3 months, (2) gives periodicity around 1.5 months, (3) shows periods around 1.6, 2.5 and 18.6-18.9 months, (4) gives periods around 1.4 and 3.4 months, (5) shows periods around 1.7 and 18.5-19 months and (6) gives periodicities around 1.8 and 2.2 months.

The observed periods of around 1.5 months for (2), 1.6 months for (3), 1.4 months for (4), 1.7 months for (5) and 1.8 months for (6) fall within the region of 10-60 days periodicity estimated by Sturrock and Scargle [16] for the SAGE and GALLEX-GNO data. Raychaudhuri [17] studied monthly solar neutrino production rate in GALLEX experiment from May 1991 to May 1992 and have found that solar neutrino production rate contains periods around $3 \pm 0.2$ months. These periods are not appreciably different from 2.5 months period obtained in (3) and 3.4 months period obtained in (4).